\documentclass{vgtc}                          
\ifpdf
  \pdfoutput=1\relax                   
  \usepackage{graphicx}                
  \DeclareGraphicsExtensions{.pdf,.png,.jpg,.jpeg} 
\else
  \usepackage{graphicx}                
  \DeclareGraphicsExtensions{.eps}     
\fi%

\graphicspath{{figures/}{pictures/}{images/}{./}} 

\usepackage{microtype}                 
\PassOptionsToPackage{warn}{textcomp}  
\usepackage{textcomp}                  
\usepackage{mathptmx}                  
\usepackage{times}                     
\usepackage{cite}                      
\usepackage{tabu}                      
\usepackage{booktabs}                  
\usepackage[table]{xcolor}
\usepackage{gensymb}

\definecolor{mycolor1}{HTML}{f0f9e8}
\definecolor{mycolor2}{HTML}{bae4bc}
\definecolor{mycolor3}{HTML}{7bccc4}

\onlineid{0}

\vgtccategory{Research}

\vgtcinsertpkg

\title{Evaluation of Text Selection Techniques in Virtual Reality Head-Mounted Displays}

\author {Wenge Xu\thanks{e-mail: wenge.xu@bcu.ac.uk}\\ %
          \parbox{1.8in}{\scriptsize \centering DMT Lab, Birmingham City University\\ Birmingham, UK} %
\and Xuanru Meng\thanks{e-mail: xuanru.meng18@student.xjtlu.edu.cn}, Kangyou Yu\thanks{e-mail: kangyou.yu18@student.xjtlu.edu.cn}\\ 
             \parbox{1.7in}{\scriptsize \centering Xi'an Jiaotong-Liverpool University\\ Suzhou, China} %
\and Sayan Sarcar\thanks{e-mail: sayan.sarcar@bcu.ac.uk}\\
             \parbox{1.8in}{\scriptsize \centering DMT Lab, Birmingham City University\\ Birmingham, UK} 
\and Hai-Ning Liang\thanks{e-mail: haining.liang@xjtlu.edu.cn (\textit{corresponding author})} \\ %
     \parbox{1.7in}{\scriptsize \centering Xi'an Jiaotong-Liverpool University\\ Suzhou, China}}

\teaser{
  \centering
  \includegraphics[width=\linewidth]{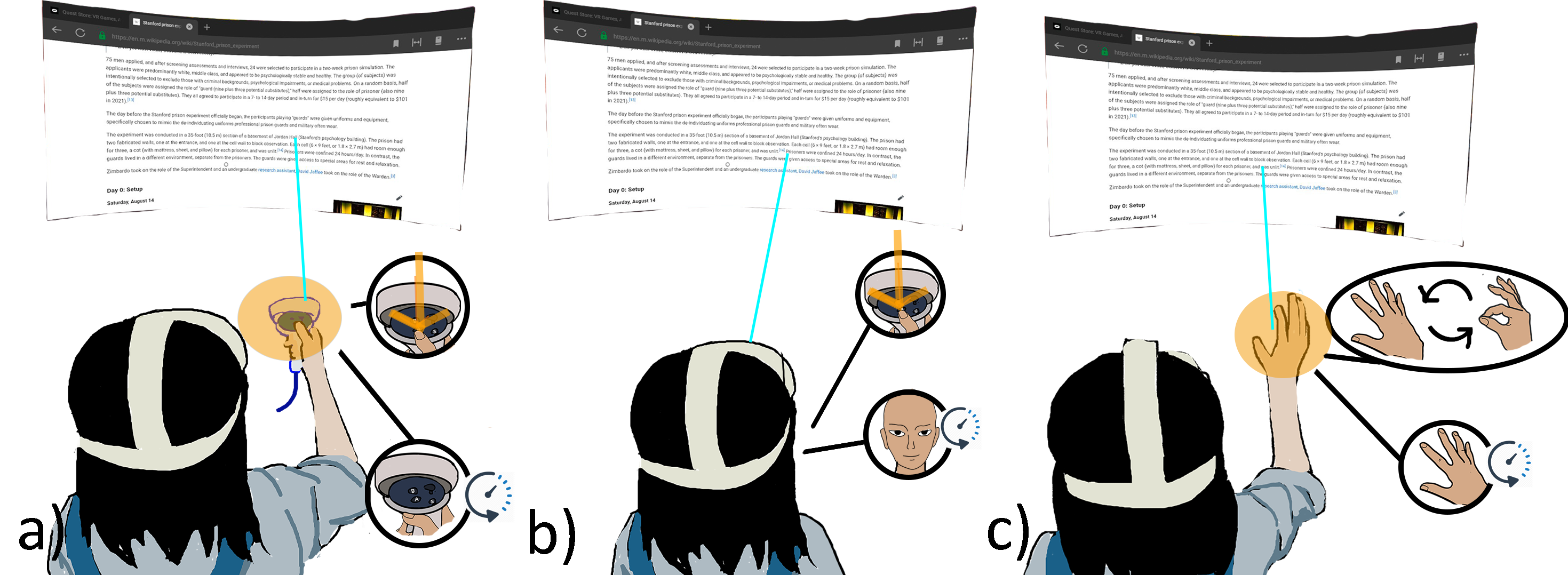}
  \caption{The six text selection techniques explored in this research, grouped by three pointing mechanisms: (a) \textit{Controller}-based: Controller+Click and Controller+Dwell, (b) \textit{Head}-based: Head+Click and Head+Dwell, and (c) \textit{Hand}-based: Hand+Pinch and Hand+Dwell.}
  \label{fig:teaser}
}

\abstract{Text selection is an essential activity in interactive systems, including virtual reality (VR) head-mounted displays (HMDs). It is useful for: sharing information across apps or platforms, highlighting and making notes while reading articles, and text editing tasks. Despite its usefulness, the space of text selection interaction is underexplored in VR HMDs. 
In this research, we performed a user study with 24 participants to investigate the performance and user preference of six text selection techniques (Controller+Dwell, Controller+Click, Head+Dwell, Head+Click, Hand+Dwell, Hand+Pinch). Results reveal that Head+Click is ranked first since it has excellent speed-accuracy performance (2nd fastest task completion speed with 3rd lowest total error rate), provides the best user experience, and produces a very low workload\textemdash followed by Controller+Click, which has the fastest speed and comparable experience with Head+Click, but much higher total error rate. Other methods can also be useful depending on the goals of the system or the users. As a first systematic evaluation of pointing$\times$selection techniques for text selection in VR, the results of this work provide a strong foundation for further research in this area of growing importance to the future of VR to help it become a more ubiquitous and pervasive platform.  
} 

\keywords{Text Selection, Virtual Reality, Pointing Methods, Selection Mechanisms, User Study}

\CCScatlist{
  \CCScatTwelve{Human-centered computing}{Human computer interaction (HCI)}{Interaction paradigms}{Virtual reality};
  \CCScatTwelve{Human-centered computing}{Human computer interaction (HCI)}{interaction techniques}{};
  \CCScatTwelve{Human-centered computing}{Interaction design}{Empirical studies in interaction design}{}
}

\begin{document}


\firstsection{Introduction}

\maketitle

Text selection is essential when reading text content such as newspapers, magazines, and academic papers to highlight important elements for later reference or to copy/cut and transfer the content to another document, application, or platform. Text selection has been well-researched for PCs (e.g., using the mouse \cite{lin_evaluating_2015,card_evaluation_1978}) and, to a lesser extent, for touchscreen devices (e.g., using a touch-based text handler in smartphones or tablets). More recently, there have been some initial explorations with augmented reality (AR) head-mounted displays (HMDs) (e.g., using a smartphone for text selection in AR HMDs \cite{darbar_exploring_2021}). However, these methods are not suitable for virtual reality (VR) HMDs because (1) traditional input devices like the mouse require a flat surface to operate on \cite{thomas_glove_2002}, which is not accessible or natural to VR users, (2) using a touchscreen device or smartphone as a dedicated input device for VR HMDs is not practical and may not lead to good performance and usability, (3) these external devices are not visible for VR users.

Today's VR HMDs enable users to chat with friends (e.g., via VR Facebook Messenger), watch movies/TV shows, read text-based content (e.g., news, blogs, papers), and even work in a virtual office (e.g., Horizon Workrooms by Facebook) \cite{grubert_office_2018,ofek_towards_2020}. These activities typically involve selecting text fragments frequently. For instance, people use text selection for (1) sharing information across platforms or applications (newspaper apps, Tripadvisor, IMDB) with their friends when exchanging instant text messages, (2) searching a movie review on IMDB before watching it, (3) making notes while reading academic papers or books, and (4) text editing \cite{darbar_exploring_2021}. Despite the increasing need for text selection tasks in VR HMDs, to our knowledge, there has been no study that has explored text selection techniques for these devices, especially based on common pointing and selection mechanisms available in current commercial VR HMDs.

There is some research on standard mid-air pointing tasks (e.g., object acquisition task \cite{7833028,9284722, yu.2021.gaze} and virtual keyboard-based text entry task \cite{selection_based_text_entry,xu_pointing_2019}). However, text selection is very different from these tasks because (1) it requires users to hold the selection mechanism until they finish hovering over the target sentence (more precise control); (2) the density of text information is always high during the task: letters and sentences tend to be close to one another; (3) the size of the letters are very small for selection compared to objects in other typical selection tasks. Nevertheless, like other mid-air pointing tasks, text selection in VR HMDs requires (1) a mechanism for the identification of the objects to be selected, and (2) some signal or command to indicate their selection \cite{Mine95virtualenvironment}. 

At present, \textit{Controller}-based pointing (see Fig. \ref{fig:teaser}a) \cite{3duserinterface,rick_survey_2010, yu.2018.targetselection} and \textit{Head}-based pointing (see Fig. \ref{fig:teaser}b) \cite{tap_dwell_gesture,xu_ringtext:_2019} are the most commonly used pointing methods in commercial VR HMDs for identifying a target object that needs to be selected. Controller-based pointing has become the dominant interaction method because it (1) offers the highest throughput and best accuracy, and is often preferred by most users \cite{controller_based_input}, (2) does not cause serious hand/arm fatigue, (3) provides rich functionalities when operating with the pointing device's built-in buttons. Head-based pointing (1) is important when controllers cannot be tracked, are not around, or has no power \cite{xu_ringtext:_2019,lu_exploration_2020}, and (2) represents a fitting alternative for people with hand mobility or stability issues. With recent developments of high accuracy and low latency hand tracking techniques, a third pointing method\textemdash \textit{Hand}-based (palm) ray-casting pointing has gained rapid attention and has been integrated into several VR HMDs (such as the Quest and HTC VIVE series). For instance, Quest 2 renders a ray cast from the point between the user’s thumb and index finger to the virtual environment to serve as a pointing mechanism (see Fig. \ref{fig:teaser}c). 

There are three primary selection mechanisms in VR HMDs to allow users to express selection\textemdash (1) \textit{Dwell}: using a dwell time on the target, (2) \textit{Click}: clicking a button on the controller, and (3) \textit{Pinch}: using one hand to perform a pinch gesture \cite{xu_dmove:_2019,pinch_dwell}. In this research, we conduct a study with 24 participants to evaluate the performance and usability of 6 text selection techniques (i.e., Controller+Dwell, Controller+Click, Head+Dwell, Head+Click, Hand+Dwell, Hand+Pinch)~\footnote{The reasons for excluding the other 3 possible combinations are provided in \autoref{Other Possible Combination that Are Not Used}.} for VR HMDs based on the 3 standard pointing methods (Controller, Head, Hand) and the 3 primary selection mechanisms (Dwell, Click, Pinch). Our results suggest that Head+Click and Controller+Click are the leading candidates where Head+Click should be considered first since it has excellent performance (2nd in speed with a relatively low total error rate), provides the best user experience, and produces a very low workload. Controller+Click has the best speed and a comparable workload as Head+Click, but because it has slightly lower user experience scores and a worse total error rate, it should be considered as a second option.

The main contributions of this work include: (1) \textit{A first evaluation} of six standard text selection techniques for VR HMDs regarding performance and user preference; and (2) a set of usage recommendations that are derived from our experiment results and observations during the experiment.

\section{Related Work}
\subsection{Text Selection in AR/VR HMDs}\label{textselectionrelatedwork}
To the best of our knowledge, text selection has only been done for AR HMDs. For instance, EYEditor \cite{ghosh_eyeditor_2020} uses a ring mouse for cursor navigation and text selection, where a button is used for placing the cursor before and after the text fragment to be selected while the selection is made via a touchpad. Lee et al. \cite{lee_one-thumb_2020} have employed a force-sensitive smartphone as their input device, where users exert a force on a thumb-sized circular button to select the desired text fragment. Similarly, Darbar et al. \cite{darbar_exploring_2021} have explored the use of a smartphone as the input mechanism for text selection in AR HMDs. They found that continuous touch is more efficient than discrete touch, spatial movement, and ray casting. This prior research focusing on AR all employed an external input device (such as a ring mouse or smartphone) for accomplishing text selection, which is not feasible or practical for commercial VR HMDs.
 
Text selection in VR differs from AR in several aspects. (1) \textit{VR is less problematic and complex}: text selection in VR is less of a problematic task than in AR, as there are no layer interference, color blending, and layout foreground-background issues \cite{xu_pointing_2019}. As such, text can be presented in a legible and readable manner to users \cite{kruijff_perceptual_2010}. (2) \textit{VR has more input possibilities to choose from}: the primary pointing methods such as \textit{Controller}-based pointing, \textit{Head}-based pointing, and \textit{Hand}-based pointing are more accessible for VR users than AR users. For instance, today's VR HMDs often come with a dedicated controller for input, while this is not always the case for AR HMDs (in fact, only the Magic Leap 1 comes with a hand-held controller). Finally, (3) \textit{VR as the future work/office space}: based on current developments of AR/VR technology, text selection in text editing and document preparation scenarios is more practical and feasible in VR for possible future office settings (e.g., VR office environments based on immersive HMDs \cite{grubert_office_2018,ofek_towards_2020}). Given the above, this study explores accessible interaction techniques that are available in commercial VR HMDs for text selection. 

Our work mainly focuses on user-driven text selection, where users have to indicate both the start and the end of the text fragment. We did not focus on automatic text selection, such as the double-click technique for word selection and the triple-click technique for the whole paragraph text selection derived from desktop computers with a mouse. Instead, we decided to explore user-driven text selection because it is a more suitable starting point for text selection studies, as it covers all types of text selection levels (e.g., character, word, sub-word \cite{darbar_exploring_2021}). On the other hand, automatic text selections are limited to word-, sentence-, or paragraph-level selection. This is in line with early text entry studies where character-level input is first explored then the swipe-/gestural-based (i.e., word-level) input followed after. 

\subsection{Mid-air Pointing-based Interaction in HMDs}
There are three main interaction approaches in VR HMDs to allow users to identify virtual objects\textemdash \textit{Controller}-based, \textit{Head}-based, and \textit{Hand}-based. Below, we present a review of these pointing methods and potential selection mechanisms that can be used to complete a mid-air interaction \cite{Mine95virtualenvironment}. 

\textit{Controller-based}. Like the mouse in desktop systems, the controller has become a standard input device for most HMDs (e.g., Quest series, HTC VIVE series, Magic Leap series) \cite{3duserinterface,rick_survey_2010,yu.2018.targetselection}. In this method, a ray is cast from the controller to the virtual environment to serve as a pointing mechanism. The end of the ray is akin to a cursor to assist in the object's identification. Users need to point to the target object by rotating or moving the controller mid-air and, once the cursor is on the desired object, the selection is often achieved by clicking a button on the controller \cite{3duserinterface}. When clicking the button becomes a problem (e.g., for very small objects or an object located in very close proximity to other objects), a dwell time can be used as the selection mechanism \cite{xu_pointing_2019}.

\textit{Head-based}. Head-based interaction has been actively studied in HMDs \cite{headinputsample,xu_dmove:_2019,3duserinterface, lu.itext, lei.2021.pointing}. It has been widely adopted as a standard way for pointing at virtual objects without using hands or hand-held pointing devices \cite{pinpoint}. Selection for head-based pointing often relies on a dwell time \cite{pinch_dwell,lu_exploration_2020,tap_dwell_gesture}, a second modality such as clicking a button on a controller \cite{pinpoint,tap_dwell_gesture}, or making a gesture using bare hands \cite{xu_dmove:_2019,pinch_dwell}. One prior study in AR that empirically compares selection methods shows that Dwell causes a lower error rate than clicking a button on a keyboard and a hand-held button device, voice (verbally confirming) \cite{6162910}. A recent study in VR also shows that Dwell leads to a significantly lower error rate than button clicking and hand gesture (i.e., pinch) \cite{pinch_dwell}. 

\textit{Hand-based}. Hand-based interaction is now possible in several VR HMDs (e.g., Quest series, HTC VIVE series). To select a nearby object \cite{Mine95virtualenvironment,yu.2018.targetselection}, users first hover their hand over the target object and then make a selection by performing a hand gesture—e.g., a pinch in the Oculus Quest 2. For a distant object (e.g., keys on a keyboard or items on a menu), users use hand-based pointing to point at the object followed by a hand gesture for selection. For instance, the Quest 2 renders a ray cast from the point between the user’s thumb and index finger to the virtual environment, and the selection of the object can be made via a pinch gesture. In scenarios when performing gestures can lead to errors, users can use dwell to prevent these errors.

Although these three mid-air interaction methods have been widely studied in HMD systems \cite{xu_pointing_2019,tap_dwell_gesture,xu_dmove:_2019,pinch_dwell}, their performance and usability for text selection tasks are underexplored. Due to the increasing importance of these tasks, understanding how these methods affect user performance and usability is crucial to the development of efficient and usable techniques for text selection. Therefore, this research aims to fill this gap and has conducted the first systematic study of these pointing mechanisms in combination with their suitable selection mechanisms for text selection in VR HMDs.

There are some other novel selection mechanisms such as eye blinking--intentionally blinking the eye(s) \cite{lu_exploration_2020,lu.itext}, voice input--verbally issuing a command \cite{voice_input}, and sound volume--increasing the sound volume \cite{hummer_text_entry} can also be used. They have been explored in the context of hands-free interaction in a recent work \cite{meng_handsfree_2022}. 

\begin{table*}[t]
  \caption{Summary of the advantages and disadvantages of the six text selection methods used in our study.}
  \label{tab:freq}
  \begin{tabular}{p{2.5cm} p{7.1cm} p{7.1cm}}
    \toprule
    Technique &Advantages&Disadvantages\\
    \midrule
    Controller+Dwell & (1) easy to control, (2) minimal hand fatigue, (3) less error-prone \cite{pinch_dwell}  & (1) not device-free, (2) slow due to dwell time and making users feel 'pushed' \cite{lu_exploration_2020}  \\
    
    Controller+Click & (1) easy to control, (2) minimal hand fatigue, (3) user familiarity (mouse-like interaction), (4) tactile feedback & (1) not device-free \\
    
    Hand+Dwell & (1) device-free, (2) less error-prone \cite{pinch_dwell}, (3) natural interaction & (1) slow due to dwell time and making users feel 'pushed' \cite{lu_exploration_2020}, (2) hand fatigue, (3) limited hand-interaction area as a result of a limited hand tracking area that can be captured via the front-facing cameras \cite{arboundary} \\
    
    Hand+Pinch & (1) device-free, (2) natural interaction & (1) hand fatigue, (2) limited hand-interaction area \cite{arboundary}  \\
    
    Head+Dwell & (1) device-free, (2) hands-free, (3) less error-prone \cite{pinch_dwell} & (1) slow due to dwell time and making users feel 'pushed' \cite{lu_exploration_2020}, (2) potential cybersickness \cite{yu_pizzatext:_2018} \\
    
    Head+Click & (1) hand mobility does not affect the selection, (2) minimal hand fatigue, (3) tactile feedback & (1) potential cybersickness \cite{yu_pizzatext:_2018}, (2) not device-free  \\
\bottomrule
\end{tabular}
\label{table:techniquesummary}
\end{table*}

\section{Evaluated Methods}
This section describes how the final six combinations from the three Pointing Methods (Controller, Head, Hand) and three Input Mechanisms (Dwell, Pinch, Click) are implemented and operationalized in our study and why Hand+Click, Head+Pinch, and Controller+Pinch are not considered. Table \ref{table:techniquesummary} summarizes the advantages and disadvantages of our six text selection methods. All six techniques were developed and run in Unity3D (Unity v2019.4.10f1) with default pointing methods and input functions provided by the Oculus Unity Plugin (OVRPlugin v1.61.0).

\subsection{Controller-based Pointing}
A ray is rendered from the controller to the virtual environment to serve as a pointing mechanism. 

\textit{Controller+Dwell (ConD)}. The user starts the selection of the desired text fragment by dwelling (staying on an area for 1s) at the beginning of the first letter and ends the selection by dwelling again at the end of the text. The dwell duration was decided through a series of pilot studies testing a range of thresholds from 400ms to 1s (from text entry studies \cite{grubert_dwell,xu_ringtext:_2019,Marco_dwell,lu.itext}). 

\textit{Controller+Click (ConC)}. The user needs to press and hold the main button in the controller at the beginning of the target letter to start the selection and then release the button after the cursor reaches the end of the target text to complete the selection. 


\subsection{Hand-based Pointing}
A ray is rendered from the point between the user’s thumb and index finger to the virtual environment to serve as a pointing mechanism. 
For this research, we upgraded the tracking rate to 60Hz to improve the hand tracking performance (i.e., lower latency for hand-based pointing and higher accuracy for pinch gesture recognition). In addition, we controlled the lighting of the physical environment to ensure the hand tracking was reliable and consistent during the study.

\textit{Hand+Dwell (HandD)}. HandD is analogous to ConD, but the pointing is done through Hand-based pointing. 

\textit{Hand+Pinch (HandP)}. The user needs to use their hand to point at the beginning of the target text, perform the “pinch and drag” gesture to start the selection, and then release the pinch gesture at the end of the target text fragment to end the selection. 


\subsection{Head-based Pointing}
A ray is extended from the HMD towards the viewing direction into the virtual environment. 

\textit{Head+Dwell (HeadD)}. HeadD is analogous to ConD, with the pointing achieved by Head-based pointing. 

\textit{Head+Click (HeadC)}. Selection is analogous to ConC, but the pointing is done through Head-based pointing. 


\subsection{Other Possible Combinations that Are Not Used}\label{Other Possible Combination that Are Not Used}
We did not include Hand+Click and Controller+Pinch because the existing literature suggests that one-handed interaction is often preferred by VR users over using two hands for interacting with objects in VR \cite{nanjappan_user-elicited_2018}. This is especially the case for tasks that can be done with one hand, as one-handed interaction is less physically and mentally demanding \cite{onehandtwohand} and it offers numerous benefits (e.g., allowing the other hand to hold other items or performing other tasks \cite{Karlson2006UnderstandingSM,one_hand_VR}). As such, this research has focused primarily on using one-hand.

Head+Pinch is excluded because: (1) our pilot study suggests that users are more likely to make errors as they frequently move their hand unintentionally outside the ideal hand interaction area of the device when their hand is used as a peripheral input modality \cite{arboundary}, (2) it leads to higher fatigue compared to Head+Click, which is already included as one of the evaluated techniques, and (3) it is disliked by users when it is used as a selection mechanism compared to Click and Dwell \cite{pinch_dwell}.

\subsection{Testbed Environment}
\begin{figure}[h]
  \centering
  \includegraphics[width=\linewidth]{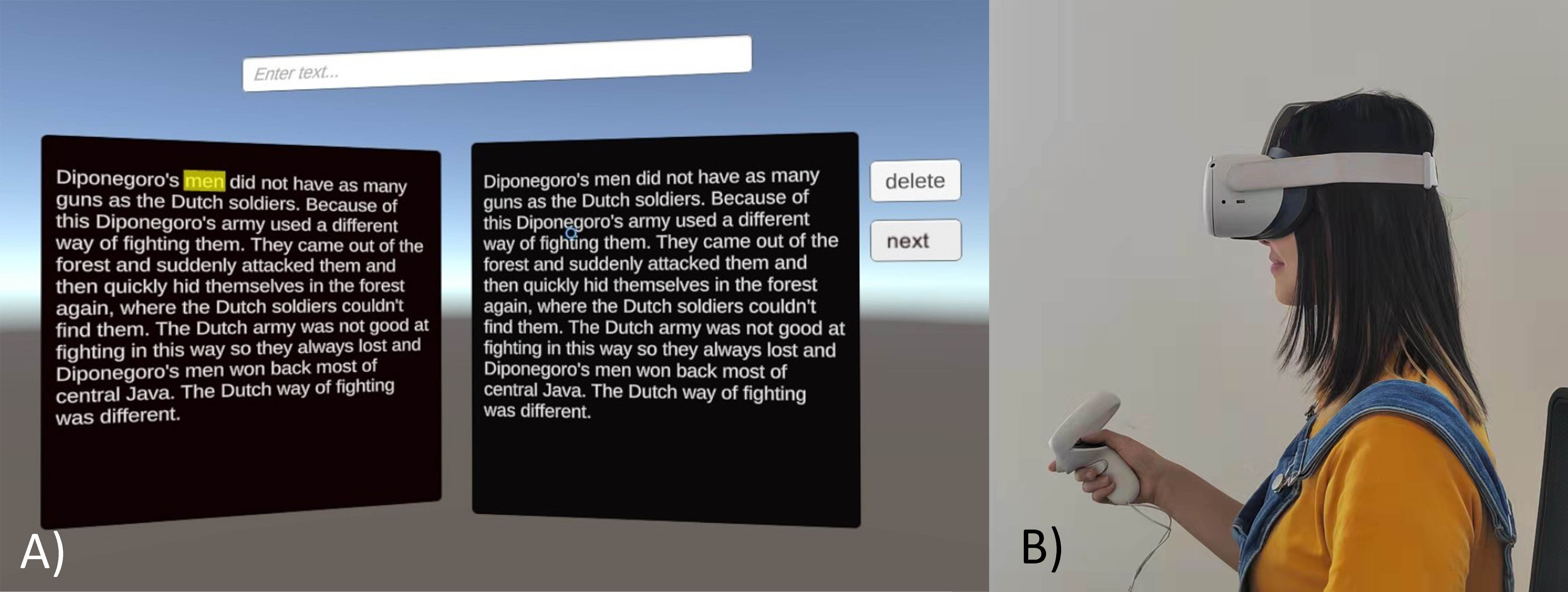}
  \caption{A) We used the same experiment environment for all conditions. An ‘instruction panel’ is located on the left side, slightly tilted towards the user. The ‘interaction panel’ is located in the center, slightly tilted towards the user. B) A picture of the experiment setting showing how a participant is performing a text selection task.}
  \label{fig:ExperimentSetup}
\end{figure}

Figure \ref{fig:ExperimentSetup} shows the test environment. An ‘instruction panel’ is located on the left side where the participant could see the text that needed to be selected. The ‘interaction panel’ is located at the center, where the participant needed to use each technique to select text fragments. In addition, two buttons are provided: ‘Delete’ for deleting the wrong selection and ‘Next’ for moving to the next trial. 

The following parameters are set based on the recommendations from previous studies and then further tested and agreed upon by 5 users from a pilot study. We controlled the length of the materials to be between 9-12 lines, with each line having around 40 characters with spaces \cite{wei_reading_2020}, in both panels. The plane is set as 2.6m which was tested and agreed by target users, 2.6m also falls in the recommended reading distance suggested by \cite{dingler_vr_2018}. For the text style, we used Sans-serif Arial with a light color \cite{dingler_vr_2018}. Angular size was set as 1.8$^{\circ}$, which was within the recommended range suggested by \cite{dingler_vr_2018}.

Visual support and feedback were provided in three ways: (1) The end of the ray is akin to a cursor, (2) changing the color of the selected text to yellow, (3) changing the color of the cursor when a selection was started/stopped, (4) a visual indicator is provided for showing dwell progress. We did not visualize the ray because users from our pilot studies suggest the cursor alone is more effective than a combination of visualizing the ray and the cursor in helping them understand where they are pointing. 

\section{Experiment}
\subsection{Participants and Apparatus}
We recruited 24 unpaid participants (12 males; 12 females) between the ages of 18-30 (M=21, SD=2.06) from a local university campus through a database of participants. They all had no issues reading the content clearly within the VR environment and no difficulties moving their arms, hand, and heads. 23 participants were right-handed. 17 participants had previous experience with VR HMDs, but only 3 were regular VR users (weekly) and none of them had interacted with the Oculus Quest 2---the device used in this experiment. Participants were seated on the office chair during the experiment and were allowed to rest their hands on the chair handles if needed (see Fig. \ref{fig:ExperimentSetup}B).

\subsection{Design}
The experiment followed a one-way within-subjects design with Interaction Techniques (HandD, HandP, HeadD, HeadC, ConD, and ConC) as the independent variable. For each condition, participants needed to complete 3 training trials (1 short, 1 medium, and 1 long text fragments) and 27 trials (9 short, 9 medium, and 9 long text fragments following a randomized order of appearance) which were randomly sampled from a corpus of standardized English reading assessment \cite{quinn_asian_2007}. Each selection target would only appear once in a specific condition. The order of the conditions was counterbalanced across participants to avoid learning effects with Sentence Lengths also randomized. Excluding the training texts, we collected 3888 trials (24 participants $\times$ 6 interaction techniques $\times$ 27 texts). Although the local area had no local COVID-19 cases for 12 months before the experiment, we cleaned and sanitized the device before and after each participant's turn and followed extra safety measures to ensure the safety of the participants and researchers (e.g., wearing a mask and staying at a safe distance). 

For assessing the text selection performance and experience, we collected the following measurements:
\begin{itemize}
\item \textit{Objective}: (1) Task-completion time: in line with the previous studies in pointing experiments, we only report task completion time where there were no errors during the trials to handle the speed-accuracy trade-off. The task completion time for each trial is defined as the time from when the cursor first hovers over the first target letter to the time they complete the correct selection. By recording in this way, the time spent on clicking the button, performing the pinch, and dwelling are included for analysis. (2) Total error rate: (the number of wrong sentences + the number of deletions) / total number of attempts, (3) Not corrected error rate: the number of wrong sentences / total number of sentences.
\item \textit{Subjective}: NASA-TLX questionnaire \cite{hart_development_1988} to measure workload, User Experience Questionnaire (UEQ) \cite{holzinger_construction_2008} to measure user experience, and comments on advantages and disadvantages of each technique plus users' ranking of the text selection techniques.
\end{itemize}

\subsection{Procedure}
Before the experiment began, participants were told the goal of the research and the experimental procedure. Then, they needed to fill out a demographic questionnaire (e.g., age, gender, and experience with VR) and signed the consent to participate in the experiment. In addition, they were told to finish the trials as fast and as accurately as they could. Error correction was allowed by using the delete button in the VR scene. 

Before each condition started, the corresponding text selection method was explained to the participants, who then had a practice session with three warm-up selections before the experiment stage (27 text selections). After each condition, participants needed to fill out a post-condition questionnaire (NASA-TLX and UEQ). They proceeded to the next condition when they felt rested to avoid any fatigue. Once they completed all conditions, they needed to complete a post-experiment questionnaire and a structured interview. The whole experiment lasted around 70 minutes for each participant.

\begin{table*}
  \caption{Performance data for each Interaction Technique among three Sentence Lengths, mean (SD). The top 3 techniques of each condition are presented by Roman numerals (I: \colorbox{mycolor1}{light green}; II: \colorbox{mycolor2}{darker light green}; and III: \colorbox{mycolor3}{blue-green}).}
  \label{tab:freq}
  \begin{tabular}{p{1.3cm} p{1cm} p{2.08cm}p{2.03cm}p{1.98cm}p{1.98cm}p{2.33cm}p{2.33cm}}
    \toprule
    Performance Metrics&Sentence Lengths&ConC&ConD&HandP&HandD&HeadC&HeadD\\
    \midrule
    Task-   & Small& \cellcolor{mycolor1}I: 1.44 (0.51)& \cellcolor{mycolor3}III: 3.08 (1.75)& 4.49 (3.79)& 3.86 (1.23) & \cellcolor{mycolor2}II: 1.69 (0.35)& 3.2 (0.87)\\
     completion & Medium & \cellcolor{mycolor1}I: 2.11 (0.65) & \cellcolor{mycolor3}III: 3.28 (0.71)& 5.20 (2.64)& 4.96 (1.65)& \cellcolor{mycolor2}II: 2.52 (0.66)& 4.38 (1.87)\\
     time & Long & \cellcolor{mycolor1}I: 2.39 (0.63)& \cellcolor{mycolor3}III: 3.71 (0.70)& 5.22 (2.25)& 5.20 (1.94)& \cellcolor{mycolor2}II: 2.92 (0.72)& 4.98 (1.47)\\
    TER  & Small & 25.7\% (18.1\%) & \cellcolor{mycolor1}I: 9.4\% (9.3\%) & 32.9\% (19.0\%) & 25.5\% (17.8\%) & \cellcolor{mycolor2}II: 11.5\% (13.8\%)& \cellcolor{mycolor3}III: 12.3\% (13.3\%)\\
         & Medium & 18.1\% (13.3\%)& \cellcolor{mycolor1}I: 8.3\% (13.0\%)& 28.2\% (17.8\%) & 22.3\%(14.4\%) & \cellcolor{mycolor3}III: 10.8\% (10.9\%)  & \cellcolor{mycolor2}II: 8.7\% (10.3\%) \\
      & Long & 20.7\% (13.4\%) & \cellcolor{mycolor1}I: 6.1\% (9.9\%) & 30.7\% (15.7\%) & 21.9\% (14.3\%) & \cellcolor{mycolor3}III: 12.0\% (13.1\%) & \cellcolor{mycolor2}II: 10.8\% (9.3\%)\\
   NCER  & Small & 1.6\% (4.9\%) & \cellcolor{mycolor2}II: 1.2\% (3.3\%) & 3.4\% (7.7\%) & 1.9\% (4.6\%) & \cellcolor{mycolor1}I: 0.0\% (0.0\%)& \cellcolor{mycolor2}II: 1.2\% (3.4\%) \\
         & Medium & \cellcolor{mycolor3}III: 1.3\% (3.9\%) & 1.6\% (4.5\%) & 3.7\% (8.2\%) & 1.4\% (3.2\%) & \cellcolor{mycolor1}I: 0.4\% (2.0\%) &  \cellcolor{mycolor2}II: 0.9\% (4.4\%) \\
      & Long & \cellcolor{mycolor1}I: 1.6\% (3.3\%) & 2.4\% (6.0\%) & 4.2\% (9.4\%) & 3.6\% (7.1\%) & \cellcolor{mycolor2}II: 2.0\% (4.8\%) & \cellcolor{mycolor3}III: 2.2\% (5.8\%) \\
\bottomrule
\end{tabular}
\label{table:performance}
\end{table*}

\subsection{Results}
We used Shapiro-Wilks tests and Q-Q plots to check if the data had a normal distribution. All tests are with two-tailed p values. Effect sizes were reported whenever feasible ($\eta_{p}^{2}$). 
\textit{Performance analysis}. For normally distributed data, we employed two-way repeated-measures ANOVAs with Interaction Techniques (six techniques) and Sentence Lengths (short, medium, long) as the within-subjects variables. For data that were not normally distributed, we processed the data through Aligned Rank Transform (ART) \cite{Jacob_ART} before using repeated-measure ANOVAs with the transformed data. Bonferroni corrections were used for all pairwise comparisons.

\textit{Experience analysis}.
For normally distributed data, we employed one-way repeated-measures ANOVAs with Interaction Techniques as the within-subjects variable. For data that were not normally distributed, we conducted non-parametric Friedman tests for the ranking. Bonferroni correction was used for pairwise comparisons and Greenhouse-Geisser adjustment was used for degrees of freedom if there were violations of sphericity.

\subsubsection{Performance}
\textit{Task Completion Time}.
To handle the speed-accuracy trade-off for task completion time, we followed the analysis process from pointing experiments where trials with errors are excluded \cite{9207831,10.1145/3355089.3356544,4142854}. In total, we collected 3888 trials (24 participants $\times$ 6 interaction techniques $\times$ 27 texts) besides training trials. To analyze selection time, we discarded trials in which participants made a wrong selection (709 error trials, 18.2$\%$), and removed outliers (65 trials, 1.7$\%$), in which the selection time was more than three standard deviations from the mean (mean $\pm$ 3std.) in each condition. In total, there were 3179 success trials used for analyzing the task completion time.

There was a statistically significant main effect of Interaction Techniques on task completion time ($F_{5,391}=145.464, p<.001, \eta_{p}^{2}=.590$). Post-hoc analysis with Bonferroni correction suggested that (1) ConC outperformed all other techniques (all $p<.001$), (2) HeadC outperformed all techniques except ConC (all $p<.001$), (3) ConD outperformed HandD, HandP, HeadD (all $p<.001$). Table \ref{table:performance} shows the mean task completion time among the 6 Interaction Techniques across 3 Sentence Lengths. 

There was a significant main effect of Sentence Lengths on Task Completion Time ($F_{2,391}=68.972, p<.001, \eta_{p}^{2}=.222$). Post-hoc analysis with Bonferroni correction suggested that (1) participants completed Small faster than Medium and Long (both $p<.001$), and (2) participants also completed Medium faster than Long ($p<.001$). We did not find any interaction effect of Sentence Lengths $\times$ Interaction Techniques ($F_{10,391}=0.981, p=.459, \eta_{p}^{2}=.019$).

\textit{Total Error Rate (TER)}. There was a statistically significant main effect of Interaction Techniques on TER ($F_{5,391}=34.391, p<.001, \eta_{p}^{2}=.270$). Post-hoc analysis with Bonferroni correction suggested that (1) ConC outperformed HandP ($p<.05$), (2) ConD outperformed ConC, HandD, and HandP (all $p<.001$), (3) HeadC outperformed ConC, HandD, and HandP (all $p<.001$), (4) HeadD outperformed ConC, HandD, HandP (all $p<.001$). There was no significant main effect of Sentence Lengths ($F_{5,391}=2.367, p=.095, \eta_{p}^{2}=.010$) and the interaction effect of Interaction Techniques $times$ Sentence Lengths ($F_{10,391}=0.406, p=.944, \eta_{p}^{2}=.008$). Details of TER data can be found in Table \ref{table:performance}. 

\textit{Not Corrected Error Rate (NCER)}. Table \ref{table:performance} presents the NCER data for each Interaction Technique. There were significant effects of Interaction Techniques ($F_{5,391}=8.217, p<.001, \eta_{p}^{2}=.073$), Sentence Lengths ($F_{5,391}=6.455, p<.01, \eta_{p}^{2}=.024$), and Interaction Techniques $times$ Sentence Lengths ($F_{10,391}=2.524, p<.01, \eta_{p}^{2}=.046$). Post-hoc analysis with Bonferroni corrections was conducted based on the interaction effect, however, we could not find any significant difference.

\begin{figure}[t]
  \centering
  \includegraphics[width=1\linewidth]{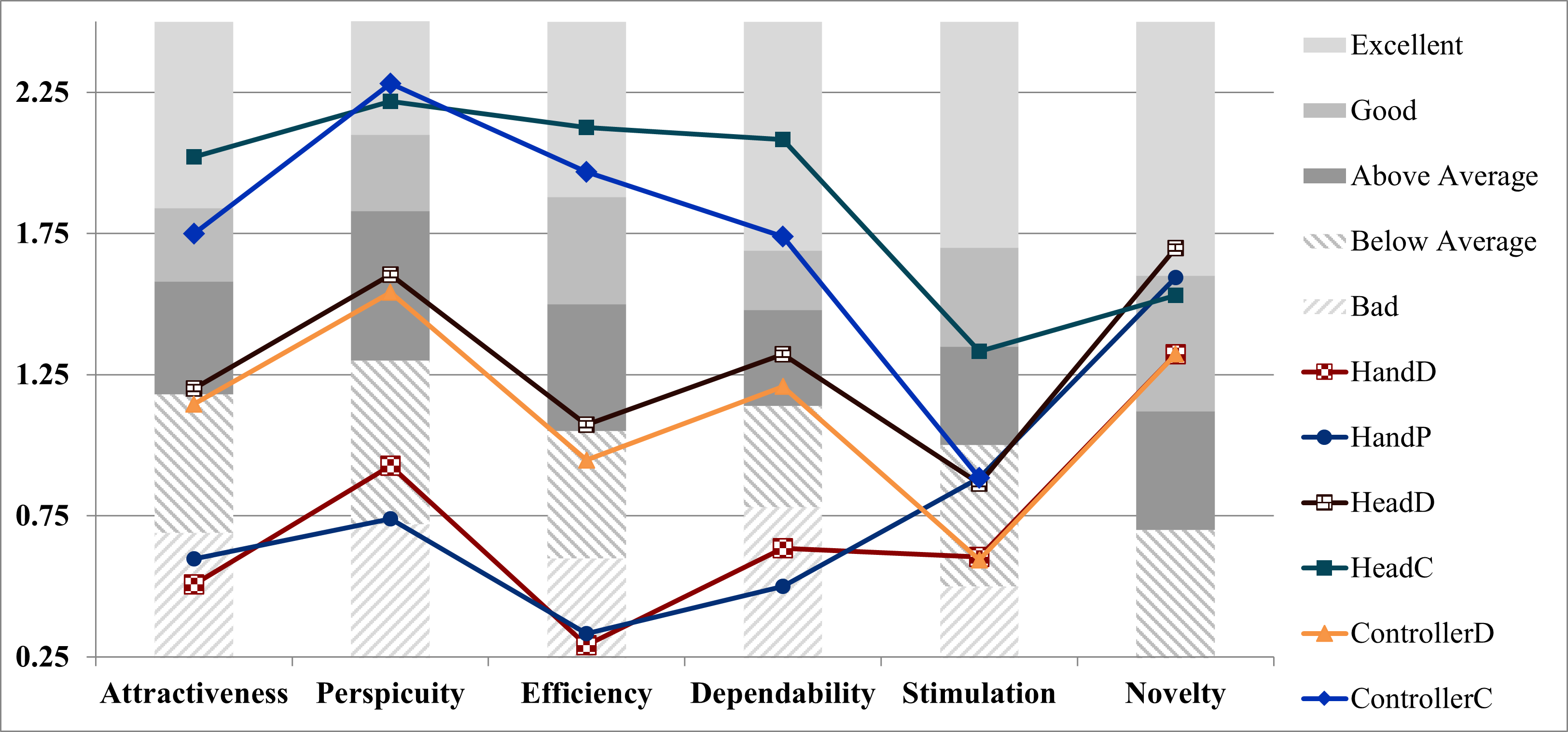}
  \caption{UEQ subscale ratings of the tested methods with respect to comparison benchmarks.}
  \label{fig:UEQbenchmark}
\end{figure}

\begin{table*}
  \caption{ANOVA test results for the UEQ subscales. Significant results where \colorbox{mycolor1}{$p<.05$} are highlighted with light green and \colorbox{mycolor2}{$p<.001$} with dark green.}
  \label{tab:freq}
  \begin{tabular}{p{2cm} p{4cm} p{11cm}}
    \toprule
    &Interaction Technique&Post-hoc\\
    \midrule
    Attractiveness &\cellcolor{mycolor2} $F_{3.307,76.066}=8.405, p<.001, \eta_{p}^{2}=.268$ & HandD$<$HeadC ($p<.01$), HandD$<$ConC ($p<.01$), HandP$<$HeadC ($p<.01$), HandP$<$ConC ($p<.05$), ConD$<$HeadC ($p<.01$) \\
   Perspicuity &\cellcolor{mycolor2} $F_{5,115}=12.631, p<.001, \eta_{p}^{2}=.354$ & HandD$<$HeadC ($p<.01$), HandD$<$ConC ($p<.001$), HandP$<$HeadC ($p<.001$), HandP$<$ConC ($p<.001$), HeadD$<$HeadC ($p<.05$), ConD$<$HeadC ($p<.05$), ConD$<$ConC ($p<.05$) \\
    Efficiency & \cellcolor{mycolor2}$F_{5,115}=15.801, p<.001, \eta_{p}^{2}=.407$ & HandD$<$HeadC ($p<.001$), HandD$<$ConC ($p<.001$), HandP$<$HeadC ($p<.001$), HandP$<$ConC ($p<.001$), HeadD$<$HeadC ($p<.05$), ConD$<$HeadC ($p<.01$) ConD$<$ConC ($p<.05$) \\
    Dependability &\cellcolor{mycolor2} $F_{3.728,85.746}=11.077, p<.001, \eta_{p}^{2}=.325$ & HandD$<$HeadC ($p<.01$), HandD$<$ConC ($p<.01$), HandP$<$HeadC ($p<.001$), HandP$<$ConC ($p<.01$), ConD$<$HeadC ($p<.01$)  \\
    Stimulation &\cellcolor{mycolor1} $F_{3.132,72.036}=2.817, p<.05, \eta_{p}^{2}=.109$ & HeadD$<$HeadC ($p<.05$), ConD$<$HeadC ($p<.01$)\\
    Novelty &\cellcolor{mycolor2} $F_{3.525,81.073}=5.835, p<.001, \eta_{p}^{2}=.202$ & ConC$<$HeadD ($p<.01$), ConC$<$HeadC ($p<.05$) \\
\bottomrule
\end{tabular}
\label{table:UEQ}
\end{table*}

\begin{figure*}[t]
  \centering
  \includegraphics[width=0.75\linewidth]{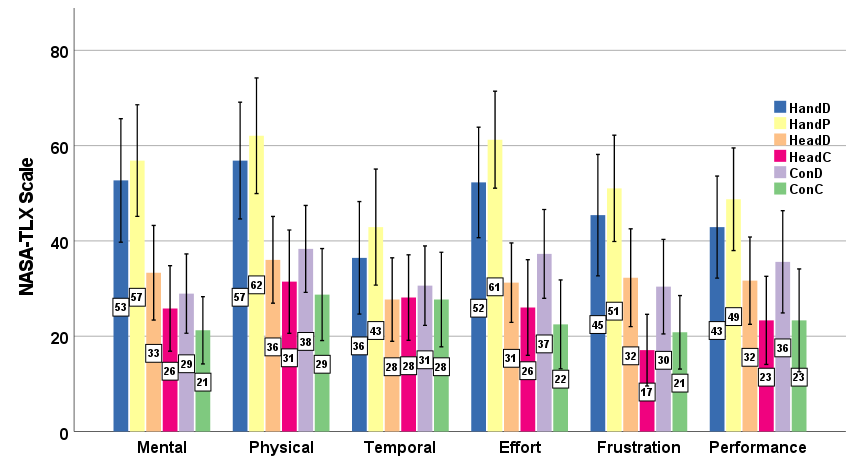}
  \caption{The mean responses for the 6 components of the NASA TLX questionnaire. Error bars indicate 95\% confidence interval.}
  \label{fig:NASA}
\end{figure*}

\subsubsection{User Preference}
UEQ. For the average UEQ scores, HeadC was rated the best (M=1.89, SD=0.77) and HandD (M=0.72, SD=1.26) the worst. ANOVA tests showed a significant main effect of Interaction Techniques ($F_{3.388,77.935}=8.622, p<.001, \eta_{p}^{2}=.273$) on the UEQ mean score. Post-hoc pairwise comparisons suggested that (1) HeadC had a significantly higher average UEQ score than HandD ($p<.01$), HandP ($p<.001$), and ConD ($p<.01$), (2) ConC had a significantly higher average UEQ score than HandD ($p<.01$) and HandP ($p<.05$). 

Regarding the UEQ subscales, ANOVA tests yielded significant main effects of Interaction Techniques on all UEQ subscales. Table \ref{table:UEQ} shows the detailed results of the ANOVA tests. Details of each UEQ subscale score can be found in Figure \ref{fig:UEQbenchmark}. 
HeadC was rated above average to excellent (mainly good to excellent except for only one participant who rated it above average), ConC was rated below average to excellent (mainly good to excellent except for one participant who rated it below average). At the same time, ConD and HeadD were comparable with the benchmark, and HandD, and HandP had worse scores than the benchmark.

\textit{Workload}. For overall task workload, ConC was rated the best (M=23.64, SD=19.32) and HandP (M=54.43, SD=20.72) the worst.
An ANOVA test showed a significant main effect of Interaction Techniques ($F_{3.168,72.871}=12.352, p<.001, \eta_{p}^{2}=.349$) on the overall task workload ratings. Post-hoc pairwise comparisons indicated that (1) users experienced less workload when use HeadC than HandD ($p<.05$) and HandP ($p<.001$), (2) users experienced less workload in ConC than HandD ($p<.01$) and HandP ($p<.001$), (3) users experienced more workload in HandP than HeadD ($p<.01$) and ConD ($p<.05$). Regarding each NASA-TLX workload subscale, ANOVA tests yielded significant main effects of Interaction Techniques among all subscales. Details of the results of the ANOVA tests can be seen in Table \ref{table:NASA} and of the workload subscales in Fig. \ref{fig:NASA}.

\begin{table*}[t]
  \caption{ANOVA test results for NASA-TLX subscales. Significant results where \colorbox{mycolor1}{$p<.05$} are shown in light green and \colorbox{mycolor2}{$p<.001$} in dark green.}
  \label{tab:freq}
  \begin{tabular}{p{1.6cm} p{4.3cm} p{11cm}}
    \toprule
    &Interaction Technique&Post-hoc\\
     \midrule
    Mental& \cellcolor{mycolor2} $F_{5,115}=11.740, p<.001, \eta_{p}^{2}=.338$ & HeadC$<$HandD ($p<.05$), ConD$<$HandD ($p<.01$), ConC$<$HandD ($p<.001$), HeadC$<$HandP ($p<.01$), ConD$<$HandP ($p<.01$), ConC$<$HandP ($p<.001$) \\
    Physical &\cellcolor{mycolor2} $F_{2.657,61.121}=7.990, p<.001, \eta_{p}^{2}=.258$ &  ConC$<$HandD ($p<.05$), HeadC$<$HandP ($p<.05$), ConC$<$HandP ($p<.01$) \\
    Temporal & \cellcolor{mycolor1} $F_{5,115}=2.919, p <.05, \eta_{p}^{2}=.113$ & N/A\\
    Effort & \cellcolor{mycolor2} $F_{2.739,63.006}=11.701, p<.001, \eta_{p}^{2}=.337$ & ConC$<$HandD ($p<.05$), HeadD$<$HandP ($p<.01$), HeadC$<$HandP ($p<.01$), ConD$<$HandP ($p<.05$), ConC$<$HandP ($p<.001$) \\
    Frustration & \cellcolor{mycolor2} $F_{3.210,73.833}=8.730, p<.001, \eta_{p}^{2}=.275$ & HeadC$<$HandD ($p<.05$), ConC$<$HandD ($p<.05$), HeadC$<$HandP ($p<.01$), ConC$<$HandP ($p<.01$) \\
    Performance &\cellcolor{mycolor2} $F_{5,115}=5.474, p<.001, \eta_{p}^{2}=.192$ & ConC$<$HandD ($p<.05$), HeadC$<$HandP ($p<.05$), ConC$<$HandP ($p<.05$) \\
    \bottomrule
\end{tabular}
\label{table:NASA}
\end{table*}

\textit{Ranking}. The ranking of conditions shows a preference for HeadC (11 ranked it first while 9 ranked it second) and ConC (11 ranked it first and 7 ranked it second) among the techniques. They were followed by ConD (10 ranked it third while 8 ranked it fourth) and HeadD (5 votes for third and 8 for fourth) as their backup. Hand-based techniques were generally the worst, being either in the fifth place (HandD with 14 votes and HandP with 5 votes) or the final option (HandP with 13 votes and HandD with 6 votes).

\subsubsection{Qualitative Feedback}
In general, most participants stated positive comments for HeadC: "\textit{simple/easy}" (N=7), "\textit{fast}" (N=6), “\textit{accurate}” (N=9), and “\textit{creative/innovative and practical}” (N=3), or simply stated, "\textit{the best}" (N=2). The only negative comments were "\textit{slightly dizzy because of the head movements}” (N=2). Similarly, most participants left positive feedback for ConC: "\textit{easy/simple and practical}" (N= 17). The reasons given were that "\textit{it just like how we interact with PC with a mouse}" (N=5). However, we also observed negative comments "\textit{tiredness/mobility of the hand lead to selection error during the selection}" (N=11), with others stating that this technique was "\textit{somewhat boring}" (N=3).

In general, participants felt positive about HeadD and ConD. They were perceived to be "\textit{fast, easy, easy to learn}" (N=11), "\textit{creative}" (N=3). The disadvantage of these methods was related to dwell-based selection "\textit{difficult to hold the pose for dwell}" (N=10), "\textit{often suffer the last sec movement which made me have to dwell again}" for Controller users (N=10), and "\textit{dizzy/eye tiredness}" for Head users (N=2).

Although HandD and HandP were perceived to be "\textit{creative/innovative and fun}" (N=8), it suffers from "\textit{tiredness}" (N=10) and sometimes are very "\textit{inaccurate (because of the difficulty in holding/performing gestures}" (N=3).

\section{Discussion}
\subsection{Task Performance}
In line with a previous study on text entry \cite{selection_based_text_entry} and supported by the bandwidth of the human muscle groups \cite{muscle_group}, we found that the default VR Interaction Techniques ConC (i.e., Controller+Click) is the best candidate regarding the task completion speed, followed by HeadC (i.e., Head+Click), and then ConD (i.e., Controller+Dwell). Despite being the fastest option, ConC led to a very high TER (21.5\%, in line with previous results in AR HMDs with a similar interaction style \cite{darbar_exploring_2021}) and NCER (1.5\%) while HeadC had a much more acceptable TER (11.4\%) and NCER (0.8\%). Consistent with a previous study that compares selection techniques \cite{pinch_dwell,ESTEVES2020102414}, Dwell-based techniques, as expected, made the least error during selection, ConD had the lowest total error rate while HeadD had the second lowest total error rate.

Hand-based interaction has been implemented in several VR HMDs (e.g., in the VIVE and Quest series). Users can now use it for system control tasks such as changing the home scene, opening/closing an application, adjusting the brightness setting, or entering texts. However, this type of interaction does not seem suitable for long-term fast-paced pointing tasks for selecting small objects. In line with the results from text entry studies (e.g., \cite{xu_pointing_2019}), we found that hand-based methods (HandD [i.e., Hand+Dwell] and HandP [i.e., Hand+Pinch]) not only had the worst task completion speed but also caused the highest number of errors. A possible reason could be due to their hand's stability; it is generally challenging to hold their hands in mid-air on a consistent and stable basis \cite{xu_pointing_2019}, even if they could use the chair handles or a table to support their hands. Another reason is that the accuracy of hand-based interaction is just not sufficient enough in today's HMDs \cite{accuracy_hand}.

In addition, we observed that the Sentence Lengths could affect the task completion time, which is as expected because participants need to move more distance across the paragraphs.

\subsection{User Experience}
Overall, HeadC and ConC are the leading options based on user experience. HeadC provided a better user experience than HandD, HandP, and ConD (e.g., average UEQ, attractiveness, perspicuity, efficiency, and dependability). In addition, HeadC was rated better than HeadD (i.e., Head+Dwell) and ConD on stimulation. ConC provided a better user experience than HandD and HandP (i.e., average UEQ, attractiveness, perspicuity, efficiency, and dependability). However, this method was considered less novel compared to HeadD and HeadC because it mimics how the mouse works for desktop and laptop computers.

When we compared these 6 text selection methods with the benchmark scores provided in \cite{UEQbenchmark}, HeadC and ConC were rated mostly good to excellent, except for the Stimulation subscale, which was only above average for HeadC and below average for ConC. HeadD and ConD were mostly rated with average scores, with hand-based methods rated mostly bad to below average. A possible explanation for why hand-based methods were rated with low user experience scores may be because they could cause hand/arm fatigue, a common issue for mid-air interaction \cite{xu_pointing_2019,xu_dmove:_2019}. Another good explanation would be that they are perceptibly slower and had more errors than controller-based and head-based approaches.

\subsection{Workload}
In line with the UEQ results, ConC and HeadC are the leading options according to the NASA-TLX workload scores. They were comparable across NASA-TLX subscales. ConC had a lower workload than HandD and HandP in all NASA-TLX subscales, except for Temporal. We also found HeadC outperformed HandP in the Mental, Performance, Effort, and Frustration scales. In addition, HeadC outperformed HandD on both Mental and Frustration scales. As mentioned by participants, the workload is high for hand-based methods because the need to hold their hands mid-air on a consistent and stable basis is difficult, which is in line with previous observations \cite{xu_pointing_2019}. It is worth noting that the Quest 2 already provides an overall stable hand tracking at a 60HZ refreshing rate. We also made sure that the light condition was consistent and in good condition during the experiment to ensure the tracking was stable. Therefore, we believe that the high workload for hand-based methods might be due to arm/hand fatigue \cite{ramos_arms_2014}, which would have made the task more complicated and unnecessarily cumbersome \cite{bowman_midair_2012}.

\subsection{Recommendations for Text Selection in VR HMDs}
The recommendations derived from our experiment are based on a set of considerations, which include performance, experience, the workload of the interaction technique, the availability of the controller, and users’ physical conditions, especially for hand/arm control. They can be divided into two groups based on the goals of the system or the users:

\textit{Performance}. If a controller is available, users could choose Head+Click since it can lead to the second fastest task completion speed and has an acceptable total error rate. If rapid neck movement is an issue for users to rotate their head for pointing, we suggest using Controller+Dwell and Controller+Click as alternative options. Controller+Dwell should be considered before Controller+Click when accuracy is needed since it has a significantly lower total error rate. If correcting errors during the session is not an issue, users can use Controller+Click. However, users should be comfortable with a relatively higher total error rate due to hand mobility limitations \cite{ramos_arms_2014}. Head+Dwell should be considered as the first option in both device-free and hands-free scenarios. As for hands-based interaction techniques (i.e., Hand+Dwell and Hand+Pinch), they should be minimized due to poor performance (in both speed and accuracy). 

\textit{User Preference}. We suggest considering Head+Click and Controller+Click as the top options when a controller is available. However, their usage should be based on the user's preference for usability and workload. If a low workload is preferred, Controller+Click can be chosen since it produces the lowest workload and provides an excellent user experience score. When a good user experience is preferred and users have no neck injuries or concerns about neck fatigue, Head+Click should be considered as the first option because it provides the best user experience score and produces a very low workload. 

Head+Dwell should be considered as the first option in device-free scenarios since it provides a better overall user experience and produces a lower workload than hand-based techniques (i.e., Hand+Dwell and Hand+Pinch). Hand+Dwell and Hand+Pinch should be minimized in rapid pointing-based text selection tasks since they generate a high overall workload and low overall user experience, with the potential to lead to the gorilla arm syndrome \cite{ramos_arms_2014}. However, they can be considered when the user has an issue with rotating their neck.

The above recommendations are derived from our study where the text selection task is assumed to be the primary task. Their inclusion in actual applications should be dependent on the context of use, target users, and availability of equipment.

\subsection{Limitations and Future Work}
One limitation of this research is that the study only consists of a single session. It will be useful to perform longer experimental sessions (e.g., 1 or 2 sessions over consecutive 4-5 days like text entry studies \cite{yu_pizzatext:_2018,xu_ringtext:_2019,wristext_2018}). Nevertheless, our study, as the first one to explore text selection in HMDs and to evaluate 6 interaction techniques from 3 common pointing methods (Controller, Head, Hand) and 3 selection mechanisms (Dwell, Click, Pinch), provides a valuable starting pointing for future studies on text selection in VR HMDs.

This work has focused primarily on user-driven text selection where the text is displayed outside users' hand reach area (i.e., far-field interaction). There are situations where (1) the text may be presented closer to users’ view and within the reach of their arms, which requires near-field interaction to manipulate the text, or (2) a user wants to select a single word or a whole sentence or paragraph akin to a double-clicking of the mouse in desktop systems (see ~\autoref{textselectionrelatedwork}). Because we want to make text selection as fast and usable as possible and in line with how users do this on their desktops, we plan to expand our work and explore new techniques and approaches that are suitable for near-field text manipulation and also can be used for automatic text selection of words, sentences, and paragraphs in the future. 

Some design considerations, such as dwell duration, visualization of the ray, and interaction distance, that have been determined via pilot studies could also be revisited to allow their fine-tuning and customization by users in actual application scenarios. In addition, we considered text selection as the primary task during the interaction with VR HMDs and future work will explore techniques to support a seamless experience on text selection to avoid the need for users to switch between techniques and interaction platforms.

We did not consider dual-hand text selection methods, given our focus on available pointing and selection mechanisms of current VR HMDs. However, they are worth exploring in the future because, when designed appropriately to minimize workload and maximize efficiency, they can extend the range of possibilities for users. For instance, researchers could consider: (1) Hand+Hand: users use their dominant hand for pointing and the other hand for selection, or vice-versa; (2) Controller+Controller: One controller is used for pointing and the other one for selection; (3) Hand+Controller: users can use their dominant hand for pointing and the controller for selection; (4) Controller+Hand: users can use a controller with their dominant hand for pointing and their non-dominant hand for selection \cite{dual-hand}. This exploration will produce additional results and help develop VR systems that are more usable and will take us one step closer to a practical immersive office of the future.

\section{Conclusion}
In this work, we have implemented six text selection techniques that resulted from the combination of three pointing methods (Controller, Head, Hand) and three selection mechanisms (Dwell, Click, Pinch) that are widely available and frequently used in commercial virtual reality (VR) head-mounted displays (HMDs). Then, we empirically evaluated these six techniques through a user study with 24 participants to assess their performance and user experience. In general, our results suggest that Head+Click and Controller+Click are the leading candidates among the 6 techniques. Head+Click should be considered as the first option since it has an excellent performance, provides the best user experience, and produces a very low workload\textemdash followed by Controller+Click, which has comparable results with Head+Click, except for having a higher total error rate. Controller+Dwell has the lowest total error rate and should be preferred if a low error rate is a plus for the user. Our results also show that current hand-based techniques should be the last possible options for rapid text selection tasks, except when users have difficulties with neck motions.

\acknowledgments{
The authors want to thank the participants who joined the study and the reviewers for their insightful comments and useful suggestions that helped improve our paper. This work was supported in part by Xi'an Jiaotong-Liverpool University (XJTLU) Key Special Fund (\#KSF-A-03), XJTLU Research Development Fund (\#RDF-17-01-54), and Future Networks Research Fund (\#FNSRFP-2021-YB-41)}.

\bibliographystyle{abbrv-doi}

\bibliography{template}

\end{document}